\documentclass{IEEEtran}
\usepackage{cite}
\usepackage{amsmath,amssymb,amsfonts}
\usepackage{algorithmic}
\usepackage{textcomp}
\usepackage{srcltx}
\usepackage{array}
\newcolumntype{P}[1]{>{\centering\arraybackslash}p{#1}}
\usepackage{float}
\usepackage{soul}
\usepackage{graphicx}  
\usepackage{dcolumn}   
\usepackage{bm}        
\usepackage{verbatim}   
\usepackage{epstopdf}
\usepackage{footnote}
\usepackage{subfigure}
\usepackage{amssymb}
\usepackage{amsmath}
\def\BibTeX{{\rm B\kern-.05em{\sc i\kern-.025em b}\kern-.08em
    T\kern-.1667em\lower.7ex\hbox{E}\kern-.125emX}}

\begin{document}

\title{Nonreciprocal-Beam Phased-Array Antennas}
\author{Reza Karimian, Shahrokh Ahmadi and Mona Zaghloul, and Sajjad Taravati
	\thanks{This paragraph of the first footnote will contain the date on 
		which you submitted your paper for review. It will also contain support 
		information, including sponsor and financial support acknowledgment.}
	\thanks{Reza Karimian, Shahrokh Ahmadiand Mona Zaghloul are with the Department of Electrical and computer engineering,
		The George Washington University, Washington D.C, USA. (e-mail: karimian@gwmail.gwu.edu). }
		\thanks{Sajjad Taravati is with the Department of Electrical and Computer Engineering, University of Toronto, Toronto, Ontario M5S 3H7, Canada. (e-mail: sajjad.taravati@utoronto.ca)
		(e-mail:sajjad.taravati@concordia.ca)
}}


\maketitle

\begin{abstract}
	This study presents a nonreciprocal-beam phased-array antenna constituted of phase-gradient patch radiators integrated with transistor-based nonreciprocal phase shifters. Such an antenna exhibits different beams for transmission and reception states. The proposed phased-array antenna provides power amplification for both transmission and reception states, which is of paramount importance in most practical applications. In addition, in contrast to the recently proposed time-modulated antennas, the proposed nonreciprocal-beam phased-array antenna introduces no undesired time harmonics and unwanted frequency conversion, requires no radio frequency bias signal. Furthermore, the nonreciprocal phased-array antenna is lightweight and is amenable to integrated circuit fabrication. The transmission and reception beam angles, the beam shapes, and the power amplification level may be easily tuned by changing the direct current (dc) bias of the transistors and phase of the passive phase shifters. Such a nonreciprocal-beam phased-array antenna is expected to find military and commercial applications.
\end{abstract}


\section{Introduction}
\label{sec:introduction}
Phased-array antennas are key elements of military radar systems, where planes and missiles are detected by steering a beam of radio waves across the sky~\cite{skrivervik1993analysis,iluz2004microstrip,mailloux2017phased}. Such versatile antennas are now widely used and have spread to modern wireless telecommunication applications. In addition to microwave and millimeter wave applications, the phased-array principle is applied to acoustics, including medical ultrasound imaging scanners, military sonar systems, and reflection seismology for gas and oil prospecting~\cite{mailloux2017phased}. Conventional phased-array antennas are restricted by the Lorentz reciprocity theorem, where the antenna is forced to introduce identical characteristics, e.g., identical beams, gains and input matchings, for the transmission and reception states.  

Recently, nonreciprocal electromagnetic and electronic systems have gained a surge of scientific interest thanks to their unique and strong capability in wave engineering and their control over the electromagnetic wave propagation. Some of the recently proposed nonreciprocal structures include nonreciprocal antennas~\cite{Volakis_TAP_2013,Taravati_LWA_2017,taravati2018space,salary2019nonreciprocal,zang2019nonreciprocal_Yagi,zang2019nonreciprocal_2,Taravati_AMA_PRApp_2020}, nonreciprocal metasurfaces~\cite{Joannopoulos_PNAS_2012,hadad2015space,Taravati_2016_NR_Nongyro,karimian2019nonreciprocal,zang2019nonreciprocal_metas,salary2019dynamically,ramaccia2019phase,taravati_PRApp_2019,wang2020theory,taravati2020_PRApp,ra2020nonreciprocal,taravati2021programmable}. Nonreciprocity can be realized by magnetic ferrite-based structures~\cite{Lax_1962,ueda2009nonreciprocal,Kodera_TMTT_04_2009,Parsa_TAP_03_2011,Volakis_TAP_2013,mannocchi2013band,Marston_patent,zang2015relay,guo2017design}, space-time modulated media~\cite{hadad2015space,Taravati_PRB_SB_2017,ramaccia2017doppler,kord2017magnet,elnaggar2018controlling,Taravati_PRAp_2018,oudich2019space,Taravati_Kishk_TAP_2019,du2019simulation,Taravati_Kishk_MicMag_2019,chegnizadeh2020non,taravati2020_PRApp}, and transistor-loaded metamaterials~\cite{kodera2011nonreciprocal,Joannopoulos_PNAS_2012,Taravati_2016_NR_Nongyro,taravati2021full,ra2020nonreciprocal,taravati2021programmable}. However, the transistor-loaded metasurfaces may present various advantages over the other nonreciprocal nonreciprocity technologies~\cite{Taravati_2016_NR_Nongyro}, i.e. less design complexity, power amplification, lack of undesired harmonics, tunability and high efficiency.

A nonreciprocal phased-array antenna is recently proposed in~\cite{zang2019nonreciprocal_2} 
by taking advantage of the asymmetric frequency-phase transition in time-modulated patch radiators. The proposed nonreciprocity mechanism in~\cite{zang2019nonreciprocal_2} is very unique and interesting, especially due to the current scientific research interest on new properties and capabilities of time modulation. However, the proposed phased-array antenna may not be suitable for many practical applications due to the following drawbacks. Firstly, the antenna in~\cite{zang2019nonreciprocal_2} introduces undesired side-band time harmonics, which interfere with adjacent channels and lead to a crowded spectrum with significant interference between adjacent channels. Secondly, the proposed nonreciprocity in the phased-array antenna in~\cite{zang2019nonreciprocal_2} is essentially accompanied with a frequency conversion which may not be required, as the forced frequency conversion possesses a small frequency conversion ratio and hence is not useful for practical applications. Thirdly, the nonreciprocity based on time modulation requires a radio frequency bias signal (in addition to a direct current (dc) bias), which represents fabrication and usage complexity. 

Conventional transmit and receive modules for active electronically scanned arrays (AESAs) utilize transmit-receive (TR) switches or circulators at the front and back of the module~\cite{Anal_Devi}. Here, we propose a “nonreciprocal-beam” phased-array antenna, where the beam shapes, transmission and reception angles as well as magnitude of the TX and RX signals can be different and electronically controlled. The AESA technique does not provide nonreciprocal-beam operation. For instance, such a full-duplex nonreciprocal-beam is extremely useful for satellite communications, where the RX signal is incoming from a country while the TX signal is transmitted to another country. Furthermore, in this scenario, the nonreciprocal-beam antenna array may be designed in a multi-band fashion to support different frequencies so that the TX and RX signals may be at different frequencies, where the transmission and reception frequencies are very close to (or far from) each other, e.g., transmission frequency of 1 GHz and reception frequency of 1.01 GHz. The conventional technique, e.g.,~\cite{mannocchi2013band,Anal_Devi} requires ferrite-based magnetic circulators and TR switches. However, as explained before, magnet-based circulators suffer from a cumbersome architecture and are not suitable for high frequency applications, and TR switches imply “half-duplex operation which is different than the full-duplex architecture of our paper. We believe that the functionality and performance of the proposed nonreciprocal-beam phased-array antenna makes it a suitable choice for modern wireless communication systems. Some of its unique features are lightweight structure, suitability for high frequency applications, controllability and programmability, arbitrary transmission and reception gains and frequencies.

\begin{figure*}
	\includegraphics[width=0.85\textwidth]{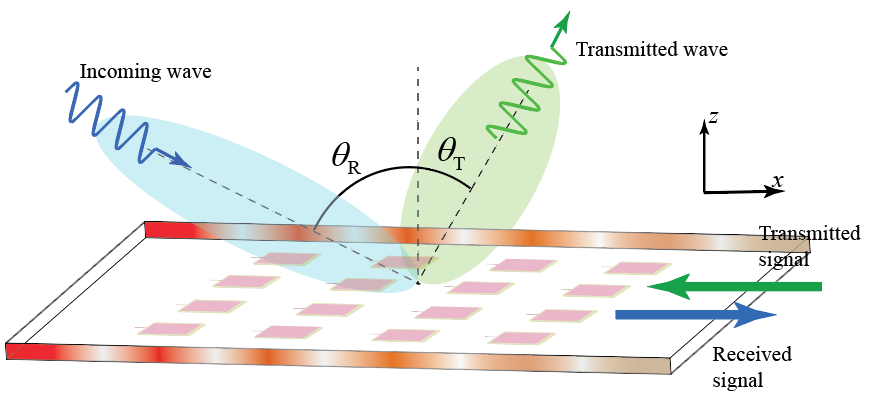} 
	\centering	\caption{Functionality of the nonreciprocal radiation pattern phased-array antenna.}
	\label{Fig:sch}
\end{figure*}

This paper proposes a nonreciprocal-beam phased-array antenna by leveraging unique properties of phase-gradient transistor-loaded nonreciprocal phase shifters. The proposed antenna introduces different beams for the transmission and reception states, which may find intriguing military and commercial applications. In contrast to the previously reported nonreciprocal/reciprocal phased-array antennas, here the antenna provides power amplification for both transmission and reception states, which is highly desired in most practical applications. Furthermore, in contrast to time-modulated antennas, the proposed nonreciprocal-beam antenna in this study introduces no undesired time harmonics and unwanted frequency conversion, and therefore, is suitable for practical scenarios. In addition, the proposed nonreciprocal-beam antenna is compatible with the planar circuit board technology. The transmission and reception beam angles as well as the power amplification can be easily tuned by adjusting the dc bias of the transistors and phase of the passive phase shifters.

The paper is structured as follows. Section~\ref{sec:concept} presents the operation principle and analytical results of the proposed nonreciprocal-beam phased-array antennas.
Then, Sec.~\ref{sec:impl} provides the details of the implementation mechanism and simulation and experimental results for the proposed transistor-based nonreciprocal-beam phased-array antenna. Finally, Sec.~\ref{sec:conc} concludes the paper.

\section{Theory}\label{sec:concept}
Figure~\ref{Fig:sch} shows the operation principle of the nonreciprocal-beam phased-array antenna. In the transmission state, the signal is launched from the input port of the antenna and radiates at the angle $\theta_\text{T}$. In contrast, in the reception state, the phased-array antenna presents the maximum reception gain for the incoming wave at the angle of reception $\theta_\text{R}$. Therefore, at a given radiation angle $\theta_0$, the phased-array antenna presents different radiation patterns for the transmission and reception states, i.e., 
\begin{equation}
E_\text{TX} (\theta)\ne E_\text{RX} (\theta),
\end{equation}
where $E_\text{TX} (\theta)$ and $E_\text{RX} (\theta)$ are the electric fields of the transmitted and received waves, respectively. As a result of this nonreciprocal-beam operation of the antenna, an incoming wave from $\theta_\text{T}$ will not be received rather is being reflected, and the transmitted wave at $\theta_\text{R}$ is supposed to be negligible. We shall stress that, as a result of this nonreciprocity, antenna may be designed in a way to acquire different radiation beam shapes, different radiation gains and different half-power beamwidths (in addition to different radiation angles) for the transmission and reception. In the next section, we provide further details on the approach for the design of such versatile antenna system. 

\begin{figure}
	\subfigure[]{\label{Fig:incl_a}
		\includegraphics[width=1\columnwidth]{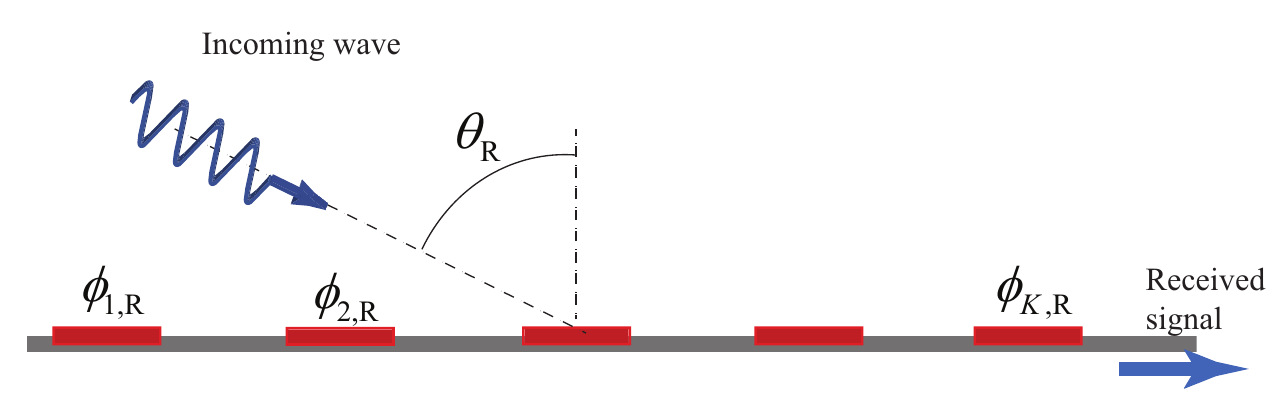}}
	\subfigure[]{\label{Fig:incl_b}
		\includegraphics[width=1\columnwidth]{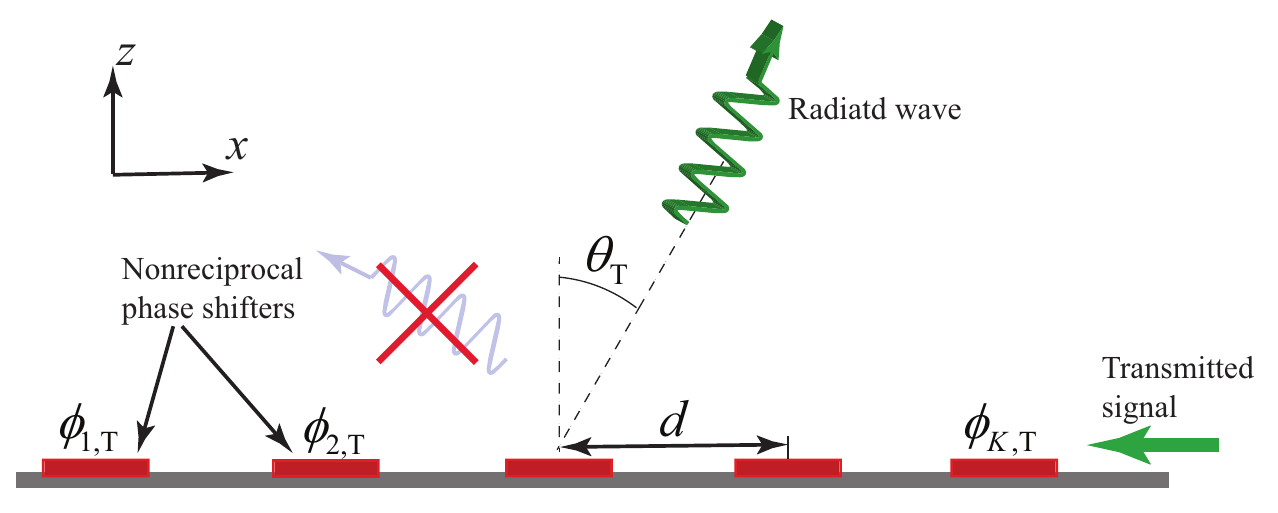}}
	\caption{Practical realization of the nonreciprocal radiation pattern phased-array antenna in Fig.~\ref{Fig:sch} using an array of phase-gradient nonreciprocal phase shifters. (a)~Reception state. (b)~Transmission state.}
	\label{Fig:incl}
\end{figure}

To realize the nonreciprocal phased-array antenna in Fig.~\ref{Fig:sch}, we consider the nonreciprocal phase-gradient phased-array antenna in Figs.~\ref{Fig:incl_a} and~\ref{Fig:incl_b}. The phased-array antenna is constituted of an array of nonreciprocal-beam microstrip patch antenna elements. The unit cells are distributed with the distance $d$, to present an arbitrary nonreciprocal radiation pattern for transmit and receive signals. For the sake of simplicity, here we assume that the antenna is uniform (no phase and amplitude difference) along the $y$ direction. Thus, the proposed antenna is supposed to introduce a nonreciprocal-beam in the $x-z$ plane as shown in Fig.~\ref{Fig:sch}. As a result, a two-dimensional beam scanning in the $x-z$ plane may be achieved by changing the characteristics of the antenna elements. However, the proposed technique in this study may be extended to a three-dimensional problem, and achieve a three-dimensional nonreciprocal-beam scanning in the both $x-y$ and $x-z$ planes.

Figure~\ref{Fig:incl_a} presents the operation of the nonreciprocal-beam phased-array antenna in the reception state, where the incoming wave under the reception angle $\theta_\text{R}$, experience gradient reception phase shifts of $\phi_{k,\text{R}}$, and is received by the antenna. In this case, the incoming waves from other directions will not be received by the antenna, rather being reflected. Fig.~\ref{Fig:incl_b} shows the operation of the phased-array antenna in the transmission state, where the transmitted wave is radiated under the transmission angle $\theta_\text{T}$. In contrast to the reception state in Fig.~\ref{Fig:incl_a}, here the radiated wave experiences the transmission phase shifts of $\phi_{k,\text{T}}$.

The array factor of the phased-array antenna may be written as
\begin{equation}\label{eq:1}
\text{AF}=I_0+I_1 e^{j\beta d \cos(\theta)}+I_2 e^{j\beta 2d \cos(\theta)}+....=\sum_{k=0}^{K-1} I_{k} e^{j\beta kd \cos(\theta)},
\end{equation}
where $d$ is the distance between two adjacent elements, $\beta$ is the wavenumber of the radiated wave, and $I_k$ is the complex current of the $k$th radiator element. By employing nonreciprocal phase shifters for each element, i.e., $I_{k,\text{TX}} \ne I_{k,\text{RX}}$. In general by assuming a linear phase progression across the antenna, we have
\begin{subequations}
	\begin{equation}\label{eq:20}
	I_{k,\text{TX}}=A_{k,\text{TX}} e^{-jk\alpha_\text{TX}},
	\end{equation}
	and
	\begin{equation}\label{eq:21}
	I_{k,\text{RX}}=A_{k,\text{RX}} e^{jk\alpha_\text{RX}}.
	\end{equation}
\end{subequations}

Here, $A_{k,\text{TX}}$ and $A_{k,\text{RX}}$ are the amplitudes of the $k$th radiator element in the transmission and reception states, respectively, and $\alpha_\text{TX}=\phi_{k,\text{T}}-\phi_{k-1,\text{T}}$ and $\alpha_\text{RX}=\phi_{k,\text{R}}-\phi_{k-1,\text{R}}$ are the phase difference between two adjacent radiator elements in the transmission and reception states, respectively. 

The array factor of the antenna reads
\begin{subequations}
	\begin{equation}\label{eq:2}
	\text{AF}_\text{TX}=\sum_{k=0}^{K-1} A_{k,\text{TX}} e^{jk (\beta d \cos(\theta)-\alpha_\text{TX})} ,
	\end{equation}
	for the transmission state, and
	\begin{equation}\label{eq:3}
	\text{AF}_\text{RX}=\sum_{k=0}^{K-1} A_{k,\text{RX}} e^{jk (\beta d \cos(\theta)+\alpha_\text{RX})} .
	\end{equation}
\end{subequations}
for the reception state.

\begin{figure}
	\subfigure[]{\label{fig:beam_sym} 
		\includegraphics[width=1\columnwidth]{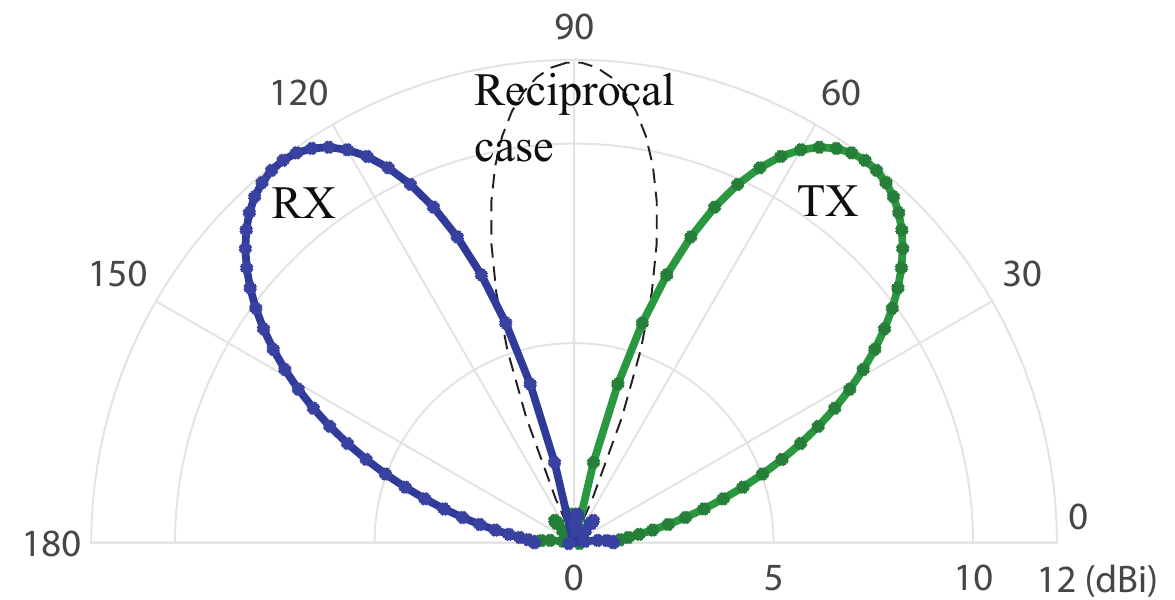}}
	\subfigure[]{\label{fig:beam_asym} 
		\includegraphics[width=1\columnwidth]{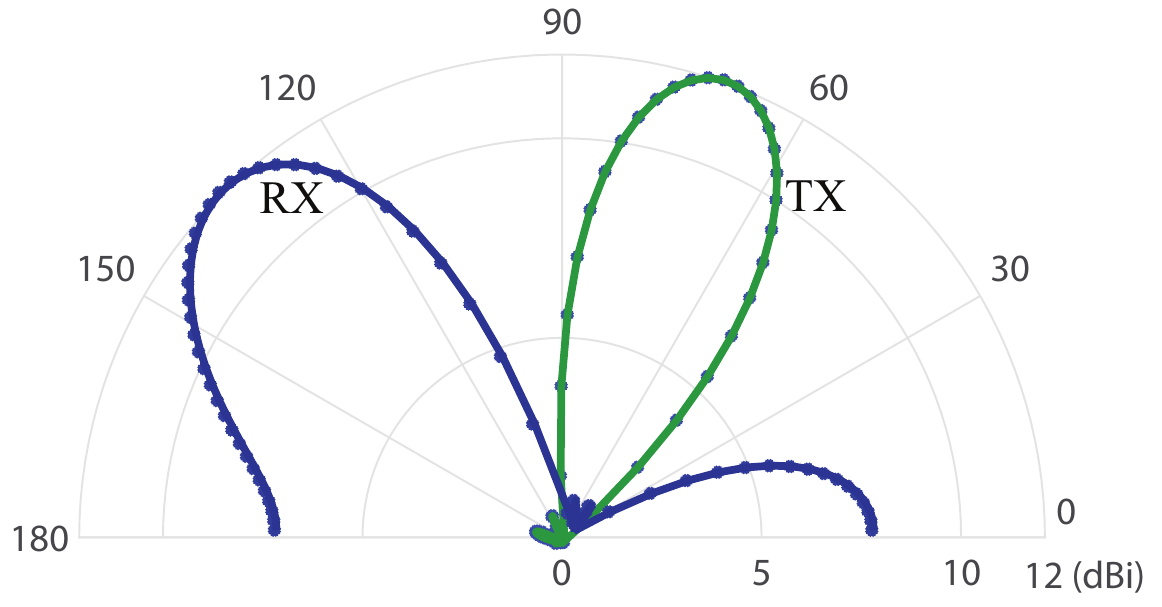}}
	\caption{Analytical results for nonreciprocal transmission and reception beams of phased-array antenna with $K=4$. (a) Symmetric beams for $\alpha_\text{RX}=\phi_{k,\text{R}}-\phi_{k-1,\text{R}}=110^\circ$ and $\alpha_\text{TX}=\phi_{k,\text{T}}-\phi_{k-1,\text{T}}=-110^\circ$. (b) Asymmetric beams for $\alpha_\text{RX}=\phi_{k,\text{R}}-\phi_{k-1,\text{R}}=130^\circ$ and $\alpha_\text{TX}=\phi_{k,\text{T}}-\phi_{k-1,\text{T}}=-60^\circ$.}
	\label{Fig:Func}
\end{figure}

Figure~\ref{fig:beam_sym} plots the analytical results using Eqs.~\eqref{eq:2} and~\eqref{eq:3} for a \textit{symmetric} nonreciprocal-beam operation of the phased-array antenna. Here, we consider four radiator elements, i.e., $K=4$, each of which introducing a nonreciprocal phase shift for the incoming and transmitted waves. In Fig.~\ref{fig:beam_sym}, an upward phase progression is considered for the reception state with $110^\circ$ phase difference between each two adjacent elements, i.e., $\alpha_\text{RX}=\phi_{k,\text{R}}-\phi_{k-1,\text{R}}=110^\circ$. However, a downward phase progression is considered for the reception state with $-110^\circ$ phase difference between each two adjacent elements, i.e., $\alpha_\text{TX}=\phi_{k,\text{T}}-\phi_{k-1,\text{T}}=-110^\circ$. As a result, the maximum gain of the transmission radiation beam occurs at $\theta_\text{T}=52.33^\circ$ with the radiation gain of 11.9 dBi. In contrast, the maximum gain of the reception radiation beam occurs at $\theta_\text{R}=127.7^\circ$ with the radiation gain of 11.9 dBi.

For some application, an asymmetric nonreciprocal-beam may be desired, where different beam shapes are achieved for the transmission and reception states. Fig.~\ref{fig:beam_asym} plots the analytical results for asymmetric nonreciprocal-beam of the phased-array antenna. Here, we consider $K=4$, $\alpha_\text{RX}=\phi_{k,\text{R}}-\phi_{k-1,\text{R}}=130^\circ$ and $\alpha_\text{TX}=\phi_{k,\text{T}}-\phi_{k-1,\text{T}}=-60^\circ$. The maximum gain of the transmission radiation beam occurs at $\theta_\text{T}=70.52^\circ$ with the radiation gain of 11.9 dBi. In contrast, the maximum gain of the reception radiation beam occurs at $\theta_\text{R}=136.24^\circ$ with the radiation gain of 11.8 dBi.

\section{Practical Implementation}\label{sec:impl}

\begin{figure}[h!]
	\includegraphics[width=0.9\columnwidth]{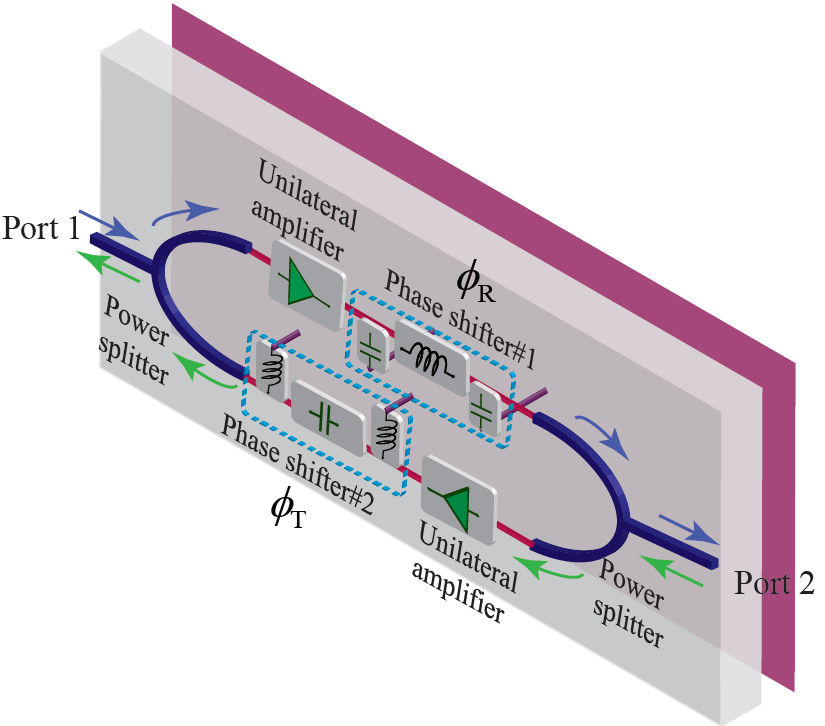} 
	\caption{Schematic representation of the transistor-based nonreciprocal phase shifter constituted of two unidirectional transistor-based amplifiers, two distributed microstrip power splitter and two passive reciprocal lumped element phase shifters~\cite{ciccognani2012active}.} 
	\label{Fig:phsh_sch}
\end{figure}

This section presents the experimental implementation of the proposed transistor-based nonreciprocal-beam phased-array antenna at microwave frequencies. To achieve nonreciprocal-beam radiator elements, we integrate microstrip patch antennas with unidirectional-transistor-based nonreciprocal phase shifters. 

Figure~\ref{Fig:phsh_sch} shows a schematic representation of the transistor-based nonreciprocal phase shifter. Such a nonreciprocal phase shifter is constituted of two unidirectional transistor-based amplifiers, two distributed microstrip power splitter and two passive reciprocal lumped element phase shifters. Two power splitters ensure full-duplex operation of the transmission and reception states, and the two unidirectional transistors ensure sufficient isolation between the transmission and reception states (i.e., $S_{12} \approx 0$) and provide power amplification in their forward operation (i.e., $S_{21} >>1$). In addition, the two (reciprocal) lumped element passive phase shifters provide the desired phase shifts for the transmission and reception states.

\begin{figure}
	\subfigure[]{\label{fig:LP} 
		\includegraphics[width=0.48\columnwidth]{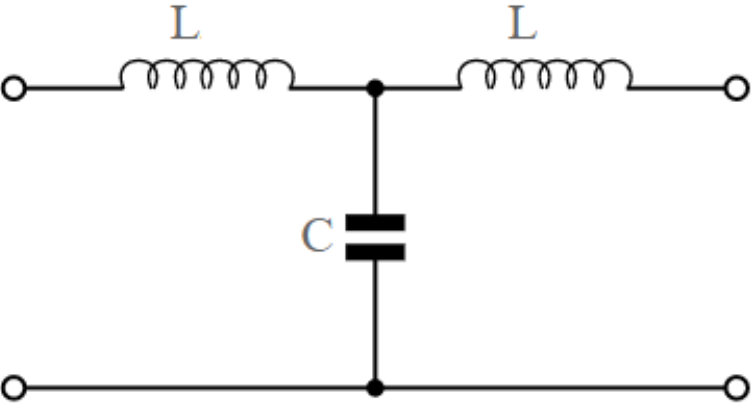}}
	\subfigure[]{\label{fig:HP} 
		\includegraphics[width=0.47\columnwidth]{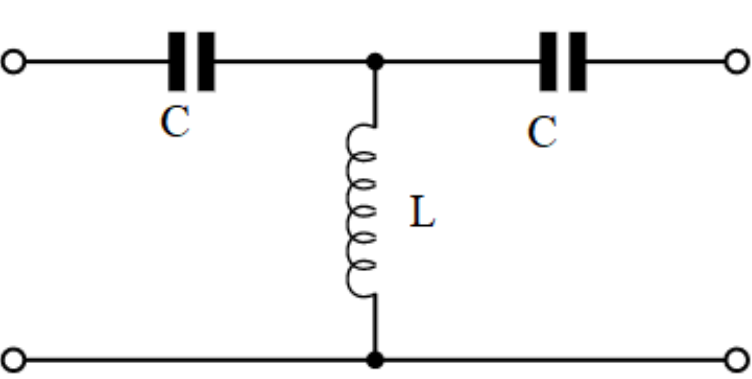}}
	\caption{Reciprocal T-type phase shifters. (a) Low pass. (b) High pass.}
	\label{Fig:Func}
\end{figure}

Figures~\ref{fig:LP} and~\ref{fig:HP} show two types of reciprocal phase shifters, i.e. a low pass T-type phase shifter and a high pass T-type phase shifter, respectively. The low pass phase shifter in Fig.~\ref{fig:LP} introduces a negative transmissive phase shift, i.e., from $-\pi$ to $0$, whereas the high pass phase shifter in Fig.~\ref{fig:HP} introduces a positive transmissive phase shift, i.e., from $0$ to $\pi$. As a result, one can achieve a complete range of desired reciprocal phase shifts from $-\pi$ to $\pi$ using these two phase shifters and construct an arbitrary nonreciprocal phase shifter using the architecture in Fig.~\ref{Fig:phsh_sch}. The inductance (L) and capacitance (C) values of the phase shifter in Figs.~\ref{fig:LP} and~\ref{fig:HP} may be found by knowing the required phase shift ($\phi$), the angular frequency $\omega$ and the characteristics $Z_0$, as	
\begin{subequations}
	\begin{equation}\label{eq:6}
	L=\dfrac{Z_0}{\omega \sin(\phi)} ,
	\end{equation}
	\begin{equation}\label{eq:7}
	C=\dfrac{\sin(\phi)}{\omega Z_0  (1-\cos(\phi)) } ,.
	\end{equation}
\end{subequations}
for the high pass phase shifter in Fig.~\ref{fig:LP}, and 
\begin{subequations}
	\begin{equation}\label{eq:6b}
	L=\dfrac{Z_0  (1-\cos(\phi))}{\omega \sin(\phi)} ,
	\end{equation}
	\begin{equation}\label{eq:7b}
	C=\dfrac{\sin(\phi)}{\omega Z_0 } ,.
	\end{equation}
\end{subequations}	
for the low pass phase shifter in Fig.~\ref{fig:HP}.

\begin{figure}[h!]
	\includegraphics[width=1\columnwidth]{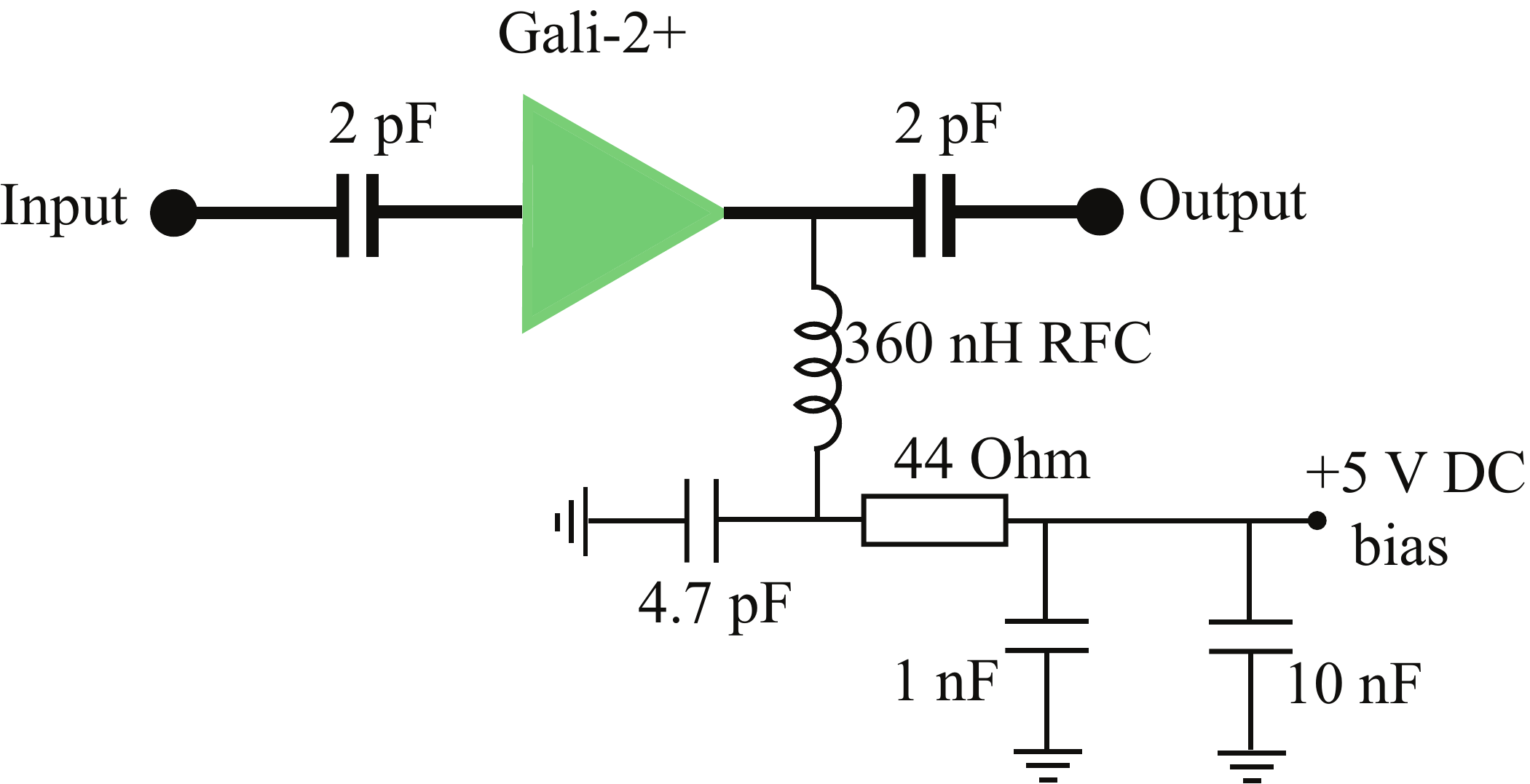}
	\caption{Circuit of the unilateral amplifier including the radio frequency path and the dc bias signal path.} 
	\label{Fig:amp}
\end{figure}

Figure~\ref{Fig:amp} shows the details of the radio frequency path and the biasing circuit of the unilateral amplifier.

\begin{figure}
	\includegraphics[width=0.9\columnwidth]{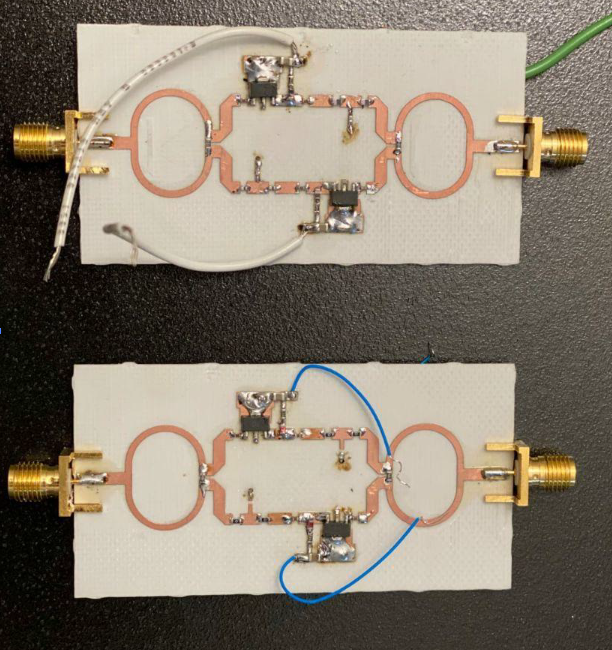} 
	\caption{Photo of the fabricated transistor-based nonreciprocal phase shifters for (top) calibration and (bottom) parametric study purposes. The fabricated prototypes are formed by Gali2+ unidirectional transistor-based amplifiers (from Mini-Circuits), two distributed microstrip Wilkinson power splitter and two passive reciprocal lumped element phase shifters.} 
	\label{Fig:phsh_photo}
\end{figure}

Figure~\ref{Fig:phsh_photo} shows an image of the two fabricated nonreciprocal phase shifters. The size of the boards is $3.5\times6.5~\text{cm}^2$. Two Gali-2 transistor amplifiers from Mini-Circuits are integrated into a microstrip structure. The two unidirectional amplifiers are placed in an unbalance scheme using two Wilkinson power splitters. The structure on the top of Fig.~\ref{Fig:phsh_photo} is fabricated for calibration purposes, where a zero phase shift for both transmission and reception channels are considered to see the effect of transistor amplifiers and their $S_{21}$ phase shift on the overall structure response. The nonreciprocal phase shifter on the bottom of Fig.~\ref{Fig:phsh_photo} is fabricated for achieving a desired nonreciprocal phase shift.

\begin{figure}
	\begin{center}
		\subfigure[]{\label{Fig:phsh_scatt}
			\includegraphics[width=0.9\columnwidth]{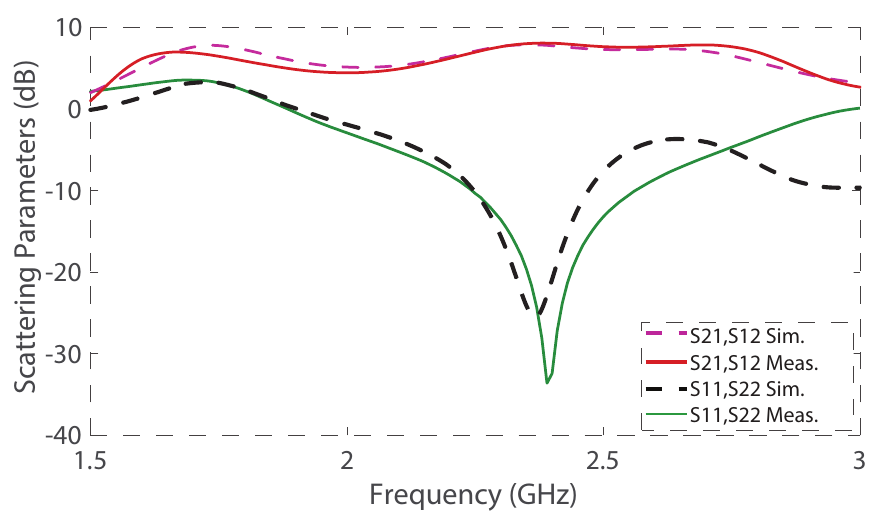}}	
		\subfigure[]{\label{fig:phsh_phase}
			\includegraphics[width=0.9\columnwidth]{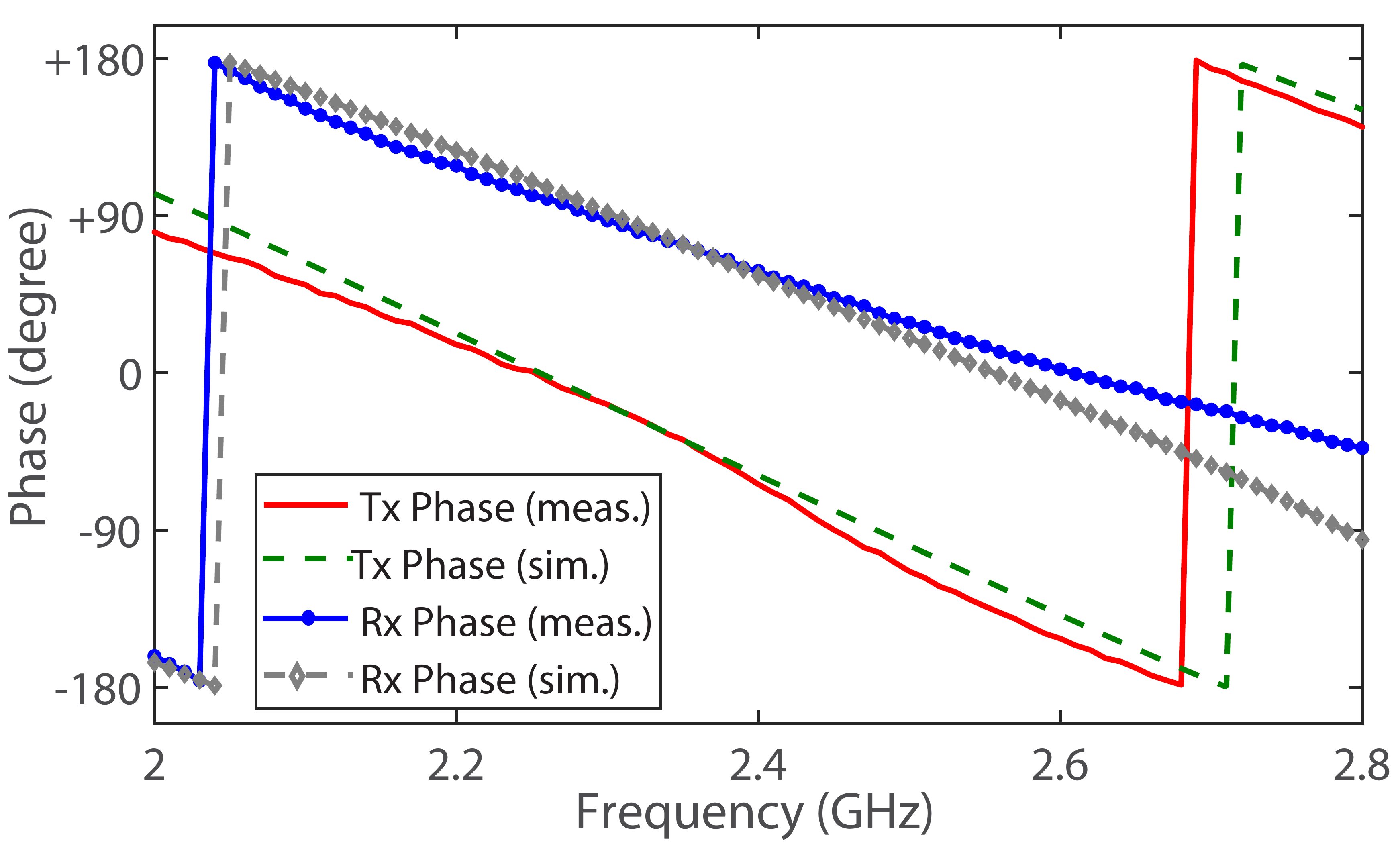}} 
		\subfigure[]{\label{fig:WNPS}
			\includegraphics[width=0.9\columnwidth]{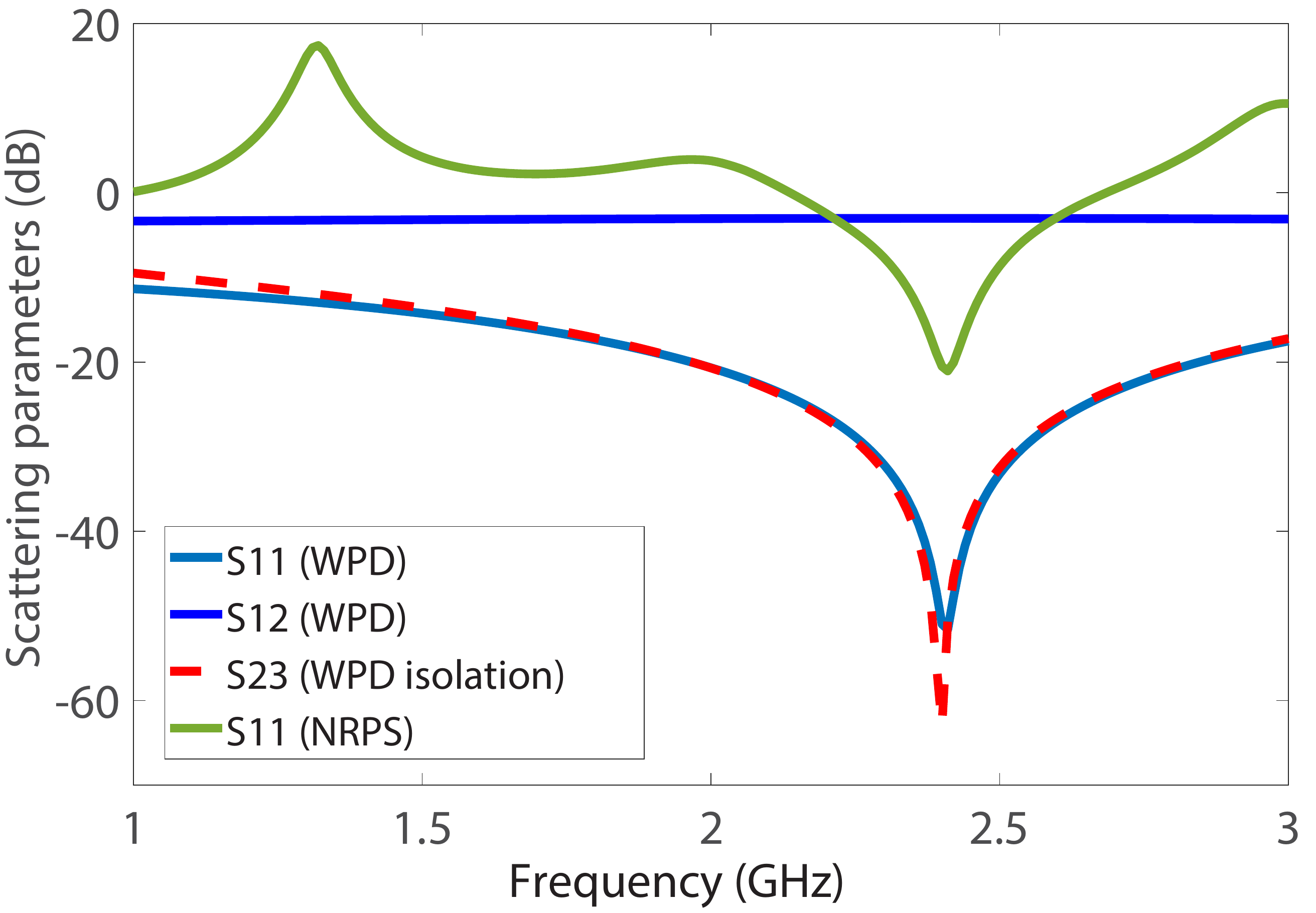}} 		
		\caption{Simulation and experimental results for the fabricated nonreciprocal phase shifter in Fig.~\ref{Fig:phsh_photo}. (a) Scattering parameters. (b) Phase response. (c) Simulation results for scattering parameters of the Wilkinson power divider and the input matching ($S_{11}$) of the nonreciprocal phase shifter.}
		\label{fig:NF}
	\end{center}
\end{figure}

Figures~\ref{Fig:phsh_scatt} and~\ref{fig:phsh_phase} plot the simulation and experimental results, respectively, for the magnitude and phase of the scattering parameters of the fabricated transistor-based nonreciprocal phase shifters, shown in the bottom of Fig.~\ref{Fig:phsh_photo}. It may be seen from this figure that the nonreciprocal phase shifter introduces more than 7 dB amplification gain ($S_{12}$ and $S_{21}$) in each direction at 2.4 GHz. The bandwidth of the nonreciprocal phase shifter is $9\%$, which is mostly limited by the bandwidth of fixed phase shifters, as amplifiers are broadband. Hence, the bandwidth of the nonreciprocal phase shifter can be increased using variable phase shifters and standard techniques for bandwidth enhancement of phase shifters~\cite{oraizi2006design,Taravati_IET_2012}. Figure and~\ref{fig:WNPS} plots the simulation results for scattering parameters of the Wilkinson power divider and the input matching ($S_{11}$) of the nonreciprocal phase shifter.

\begin{figure}
	\includegraphics[width=0.45\textwidth]{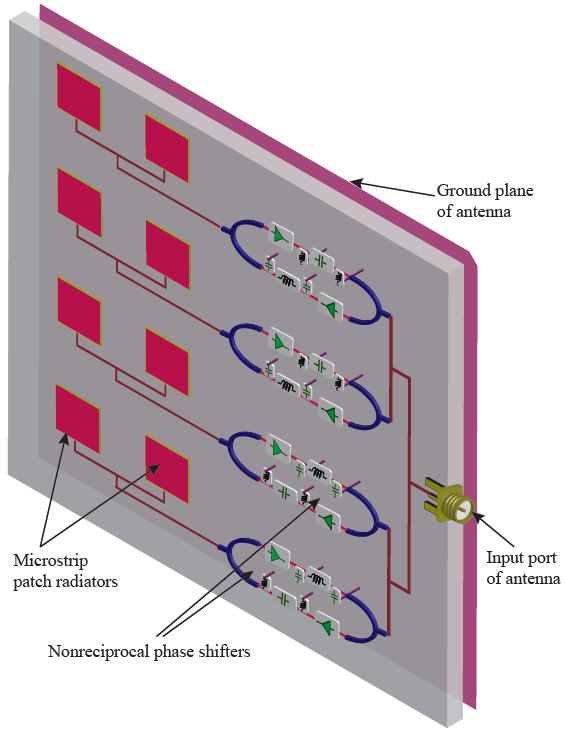}
	\centering	\caption{Schematic representation of the fabricated nonreciprocal-beam phased-array antenna composed of four nonreciprocal phase shifters and a $4\times2$ antenna array.} 
	\label{Fig:fabr_sch}
\end{figure}

Figure~\ref{Fig:fabr_sch} depicts a perspective of the designed nonreciprocal-beam phased-array antenna. The designed prototype is composed of an array of $4\times2$ microstrip patch elements and four nonreciprocal phase shifters. The antenna is designed, based on the implementation scenario in Fig.~\ref{Fig:incl}, at frequency $f=2.4$~GHz using eight microstrip patches distributed with the distance of $d=\lambda/2$. Table~\ref{tab:table0} lists the details of the eight lumped element T-junction phase shifters that are required to form the four phase-gradient nonreciprocal phase shifters.

\begin{table}
	\centering
	\caption{Inductance and capacitance values for the lumped-element phase shifters of the phase-gradient phase shifters, for $f=\omega/2\pi =2.4$ Ghz and $Z_0=50$ Ohm.}
	\label{tab:table0}
	\begin{tabular}{ |P{0.5cm}| |P{1.3cm}|P{1.4cm} |P{1cm}|P{1cm}|  }
		\hline
		& Phase shift, $\phi$ (degree)& Type& L (nH)& C (pF)\\
		\hline 	\hline
		%
		%
		1& -172   & Low pass T   &  5.1 &  1 \\\hline
		2& 170 & High pass T & 0.78  & 0.62 \\\hline
		3&  -60   &  Low pass T    &  1.7  & 0.15 \\\hline
		4& 58    & High pass T& 1.4& 1.3 \\\hline
		5& 49 &High pass T  & 1.9 & 1.8 \\\hline
		6&  -44     &Low pass T  & 0.33 &  0.18\\\hline
		7& 161  &High pass T &0.56 &   0.68  	\\\hline
		8& -151  & Low pass T& 4.9 &  0.82
		\\\hline	
	\end{tabular}
\end{table}

\begin{figure}
	\begin{center}
		\includegraphics[width=0.8\columnwidth]{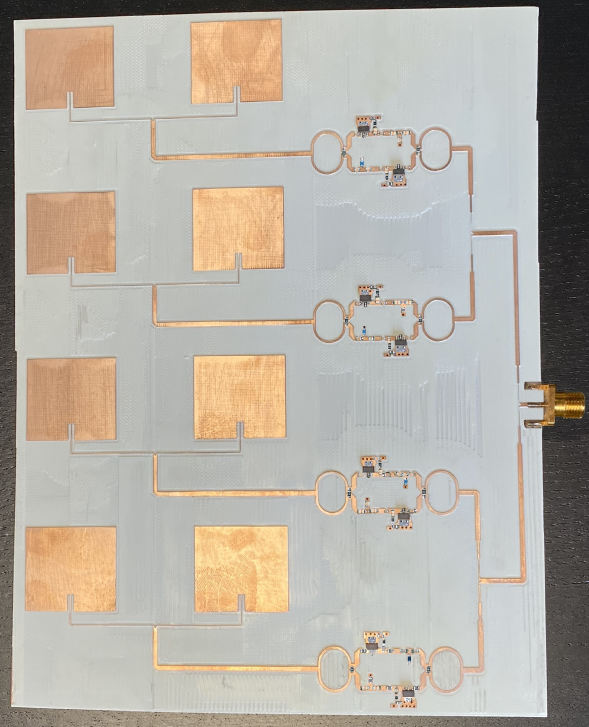}
		\caption{An image of the fabricated nonreciprocal-beam phased-array antenna prototype.} 
		\label{Fig:fabr}
	\end{center}
\end{figure}

\begin{figure}
	\subfigure[]{\label{fig:beam1} 
		\includegraphics[width=0.9\columnwidth]{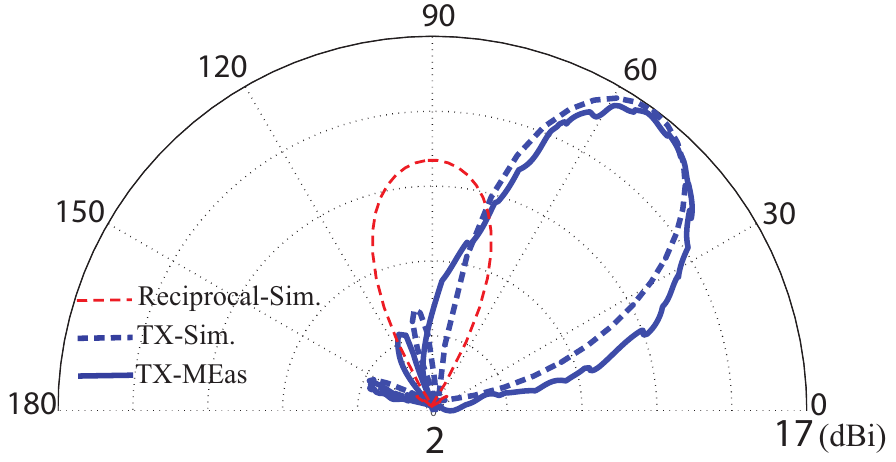}}
	\subfigure[]{\label{fig:beam2} 
		\includegraphics[width=0.9\columnwidth]{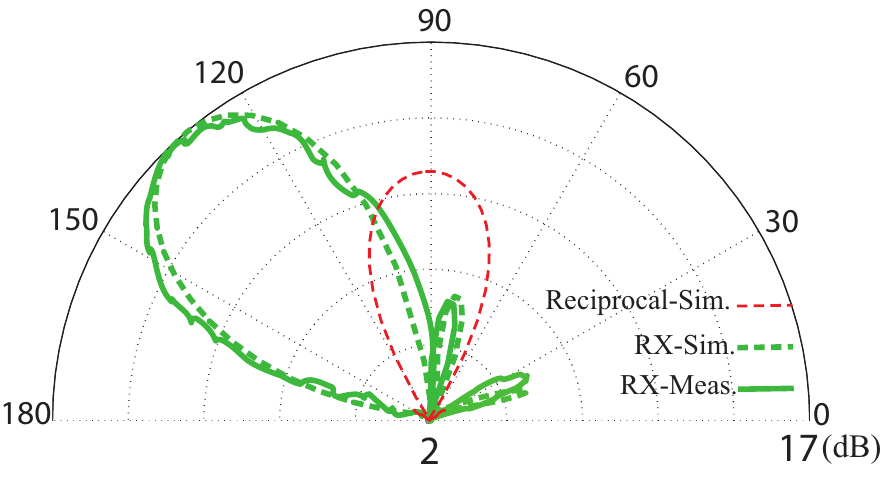}} 
	\caption{Experimental and simulation results for the designed nonreciprocal-beam phased-array antenna based on the implementation scenario in Fig.~\ref{Fig:fabr_sch}, with $f=2.4$~GHz, $K=4$, $d=\lambda/2$, $\theta_\text{T}=52.33^\circ$ and $\theta_\text{R}=127.7^\circ$. (a) Transmission (TX) radiation beam. (b) Reception (RX) radiation beam.
	} 
	\label{Fig:beams}
\end{figure}

Figure~\ref{Fig:fabr} shows an image of the fabricated nonreciprocal-beam phased-array antenna. The antenna size is $19\times25~\text{cm}^2$. Here, an upward phase progression is considered for the reception state with $110^\circ$ phase difference between each two adjacent elements, i.e., $\alpha_\text{RX}=\phi_{k,\text{R}}-\phi_{k-1,\text{R}}=110^\circ$, and a downward phase progression is considered for the reception state with $-110^\circ$ phase difference between each two adjacent elements, i.e., $\alpha_\text{TX}=\phi_{k,\text{T}}-\phi_{k-1,\text{T}}=-110^\circ$. The measurement of the TX and RX beams that are plotted in Figs.~\ref{fig:beam1} and ~\ref{fig:beam2} is accomplished using a vector network analyzer (VNA), a reference horn antenna and the fabricated nonreciprocal antenna array. The port-1 and port-2 of the VNA are connected to the input port of the fabricated antenna array and the input port of the reference horn antenna, respectively. Hence, by rotating the reference horn antenna from 0 to 180 degrees (by steps of 5 degrees), the measured the S21 and S12 provide the TX and RX patterns of the fabricated nonreciprocal-beam array, respectively. It should be noted that the antenna patterns in Figs.~\ref{fig:beam1} and ~\ref{fig:beam2} are plotted with reference to 2 dBi. These two plots show the nonreciprocal-beam of the metasurface, as well as more than 7 dB power amplification. As a result, given 10 dBi gain of the RX and TX beams, and 7 dB power gain of nonreciprocal active phase shifters, more than 17 dBi gain for TX and RX paths is achieved. Since the active phase shifters are part of this active antenna array and integrated into the antenna structure, we have plotted the pattern of the entire antenna and did not separate the gain provided by the active phase shifters.

Figure~\ref{fig:beam1} plots the full-wave simulation and experimental results of the fabricated prototype for the transmission state. The maximum gain of the transmission radiation beam occurs at $\theta_\text{T}=52.33^\circ$ with the radiation gain of 17 dBi. The simulation of the nonreciprocal antenna is carried out using CST Microwave Studio, and simulation of the nonreciprocal phase shifter is accomplished using Advanced Design System (ADS) commercial software. 

Figure~\ref{fig:beam2} plots the full-wave simulation and experimental results of the fabricated prototype for the reception state. In contrast to the transmission state in Fig.~\ref{fig:beam1}, here the maximum gain of the reception radiation beam occurs at $\theta_\text{R}=127.7^\circ$ with the radiation gain of 17 dBi.

Figure~\ref{fig:beam1} and~\ref{fig:beam2} show that more than 15 dB isolation between the transmission and reception beams, at $\theta_\text{T}=52.33^\circ$ and $\theta_\text{R}=127.7^\circ$, is achieved. Furthermore, it may be seen from Figs.~\ref{fig:beam1} and~\ref{fig:beam2} that more than 5 dB power gain is achieved compared to the conventional reciprocal phased-array antenna. Such a significant power amplification is due to the amplifications of the transistor amplifiers in both transmission and reception states. Each amplifier is supplied by a dc voltage of 3.8V and dc current of 43mA. As a result, the power consumption of the entire array which includes eight amplifiers is 1.3 Watt. It should be noted that, in this particular prototype example, we utilized a nonreciprocal phase shifter for each row (each row includes two horizontal patches), which is suitable for a two-dimensional nonreciprocal beam scanning. However, one may assume a three-dimensional nonreciprocal beam scanning where a nonreciprocal phase shifter is connected to each individual patch element. In this case, more leverage on beam scanning is achieved at the cost of increased power consumption of the entire array because of more utilized amplifiers.

\begin{figure}
	\includegraphics[width=1\columnwidth]{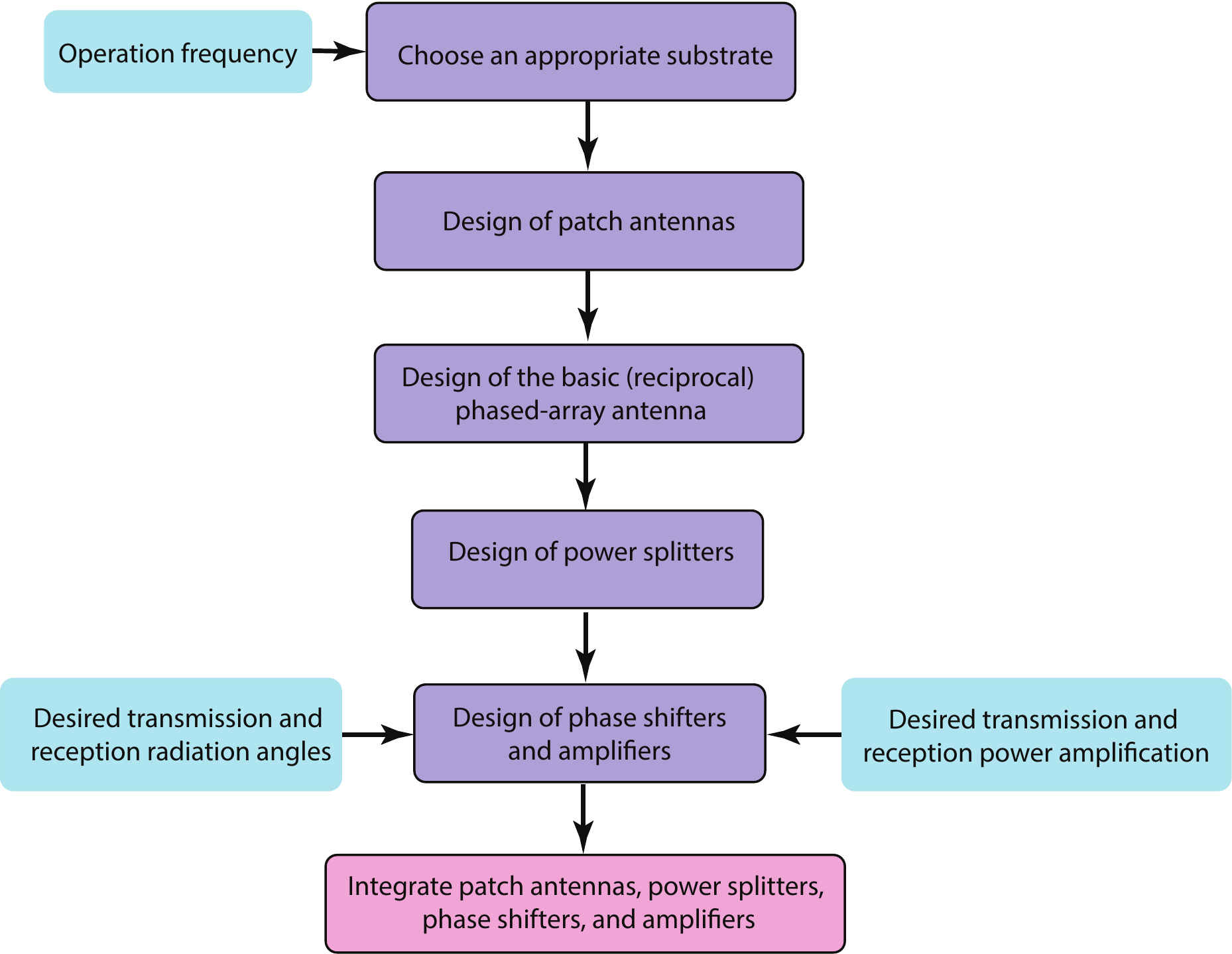} 
	\caption{Design procedure of the transistor-based nonreciprocal-beam phased-array antenna.} 
	\label{Fig:proc}
\end{figure}

Figure~\ref{Fig:proc} describes the design procedure of the transistor-based nonreciprocal phased-array antenna. Table~\ref{tab:table1} compares different features of the proposed nonreciprocal-beam phased-array antenna with of other recently proposed nonreciprocal phased-array antennas and metasurfaces. One of the main advantages of the proposed active phase array antenna is its programmability. In practice, the functionality of the antenna, e.g., the TX/RX isolation, gain, radiation angles, and stability of the active nonreciprocal phase shifters, can be programmed and controlled through a field-programmable gate array (FPGA).

\begin{table*}
	\centering
	\caption{Comparison of the recently proposed nonreciprocal phased-array antennas and metasurfaces.}
	\label{tab:table1}
	\begin{tabular}{ |P{2cm}| |P{2.6cm}|P{2cm} |P{2.5cm}|P{2.5cm}|  }
		\hline
		& Nonreciprocal transistor-based metasurface~\cite{Taravati_2016_NR_Nongyro}& Nonreciprocal-beam time-modulated metasurface~\cite{taravati2020_PRApp}& Nonreciprocal-beam time-modulated antenna~\cite{zang2019nonreciprocal_2}& This work (Nonreciprocal-beam transistor-based antenna)\\
		\hline 	\hline
		%
		%
		Full-duplex beam-steering& No & Yes & Yes & Yes\\\hline
		RF bias signal& No need& Needed & Needed & No need \\\hline
		Producing undesired time harmonics&  No &     Yes &  Yes & No\\
		\hline
		Tunable& Yes  &Yes&Yes& Yes \\\hline
		RX and TX power amplification& Only one-way amplification & No &No& Yes \\\hline
		Integrable with circuit technology& Yes    &Yes&Yes& Yes 
		\\\hline
		Advantages over conventional techniques&  Magnetless, lightweight &Magnetless, linear, lightweight &Magnetless, linear and lightweight & Magnetless, power efficient  
		\\\hline	
	\end{tabular}
\end{table*}

An important practical point is that the structure (power dividers, and other components) should be designed in a way to ensure that the required isolation between the transmission and reception is achieved for a particular transmitted power level and received power level. Otherwise, if the isolation of the power dividers is not sufficient, part of the transmitted power will leak to the reception path. Another critical point is the noise figure of the array in the reception state. The noise figure of each nonreciprocal phase shifter is equal to the noise figure of one unilateral amplifier plus 3 dB noise figure of the power divider and noise figure of reciprocal phase shifter. In the fabricated prototype, the total noise figure is equal to 8.9 dB, including 3.07 dB for power divider, 4.76 dB for Gali2+ unilateral amplifier and 1.07 dB for reciprocal lumped element phase shifter.

\section{Conclusions}\label{sec:conc}
We have presented a nonreciprocal-beam phased-array antenna by taking advantage of unique properties of transistor-based nonreciprocal phase shifters. The proposed antenna exhibits different beams for transmission and reception states. Different from the previously reported phased-array antennas, the proposed phased-array antenna in this study provides power amplification for both transmission and reception states, which is desired in practical applications. The experimental results show that more than 15 dB isolation between the transmission and reception beams is achieved, and more than 5 dB power gain is achieved compared to the conventional reciprocal phased-array antenna due to the power amplification by unidirectional transistors. In addition, in contrast to recently proposed time-modulated antennas, the proposed nonreciprocal-beam phased-array antenna introduces no undesired time harmonics and unwanted frequency conversion, and hence, is suitable for practical applications. The proposed nonreciprocal-beam antenna is compatible with the integrated circuit technology. Furthermore, the transmission and reception beam angles as well as the power amplification may be tuned through the dc bias of the transistors and phase shifts of the passive phase shifters. The proposed phased-array antenna is highly efficient and is expected to find various military and commercial applications. Furthermore, the efficiency of the proposed antenna, e.g., the frequency bandwidth and size, may be improved by using previously reported antenna and power splitter engineering techniques~\cite{tong2000broad,wong2001broad,rodenbeck2005ultra,Taravati_IET_2012}.

\bibliographystyle{IEEEtran}
\bibliography{Taravati_Reference}

\end{document}